\documentclass[letter]{aa}
\usepackage[varg]{txfonts}

\usepackage{hyperref}
\usepackage{natbib}
\bibpunct{(}{)}{;}{a}{}{,} 

\def\Msun{{\rm M_{\odot}}}

\newcommand{\ergs}{\ensuremath{\,\mathrm{erg}\,\mathrm{s}^{-1}}}

\usepackage{color}

\begin{document}

\title{SN 2022jli:  The ultraluminous birth of a low-mass X-ray binary}

\author
{Andrew King\inst{1, 2, 3}
\and
Jean--Pierre Lasota\inst{4, 5}}

\offprints{J.-P. Lasota \email{lasota@iap.fr}}

\institute{Astronomical Institute Anton Pannekoek, University of Amsterdam, Science Park 904, 1098 XH Amsterdam, Netherlands
\and Astrophysics Group, School of Physics \& Astronomy, University of Leicester, Leicester LE1 7RH, UK
\and Leiden Observatory, Leiden University, Niels Bohrweg 2, NL-2333 CA Leiden, Netherlands
\and Institut d'Astrophysique de Paris, CNRS et Sorbonne Universit\'e, UMR 7095, 98bis Bd Arago, 75014 Paris, France 
\and Nicolaus Copernicus Astronomical Center, Polish Academy of Sciences, ul. Bartycka 18, 00-716 Warsaw, Poland    
}

\date{Received/Accepted}


\abstract{
Observations show that the 12.4~d binary system descending from the recent supernova SN 2022jli
closely fits hypotheses of how low-mass X-ray binaries form, but requires an apparently super-Eddington accretion luminosity from the accreting component. We show that this agrees very well with the type of accretion-induced beaming found in ultraluminous X-ray sources, as recently strongly confirmed by X-ray polarimetry of the X-ray binary Cyg X-3. Beaming in the SN2022jli binary system occurs because of the very high mass-transfer rate induced by the violent effect of the supernova on the binary geometry. This
explains the very soft nature of the accretion luminosity, its distinctive periodic light curve, and its luminosity decay on a $\sim 250$~day timescale. A test of this picture is that the system's orbital period should increase on a $10^5$~year timescale.}


\maketitle

\nolinenumbers

\section{Introduction}

In a recent paper, \citet[hereafter C24]{Chen0124}  report the appearance of
a new Fermi--LAT gamma-ray source consistent in time and position with the supernova (SN) SN 2022jli. In addition to the nuclear-powered gamma-ray emission from the SN, C24 find 12.4 d periodic luminosity
variations of the 
UVOIR emission as the source decays on a $\sim 250$~d timescale, and narrow H$\alpha$ velocity shifts with a similar 12.4~d period. 

C24 point out that all of this evidence is strongly consistent with the SN explosion of a massive star in a binary system with a lower-mass companion. This companion has remained bound despite the explosion, because the {anisotropic} back-reaction of mass loss in the SN (the `kick') is suitably directed, as long suggested in models of X-ray binary formation \citep{Flannery0275}. 

The H$\alpha$ emission presumably arises as hydrogen stripped from the companion star accretes onto the compact object (black hole or neutron star) formed in the SN. Therefore, SN 2022jli offers a direct view of the birth event of a system that may eventually become a low-mass X-ray binary.

C24 note at least one potential problem with this picture, in that the apparent accretion luminosity of the compact SN remnant is $ \sim 10^{42}~{\rm erg\, s^{-1}}$, far exceeding the Eddington luminosity of
\begin{equation}
L_{\rm Edd} \simeq 1.3\times 10^{39}m_{10}\ergs
\label{ledd}
\end{equation}
for hydrogen-rich accretion, where $m_{10}$ is the accretor mass $M$ in units of $10\Msun$. 

We consider this 
problem further in this Letter. This leads us to a comprehensive picture of the system, with suggestions for further observations that should offer a stringent test of this picture.

\section{Ultraluminous X-ray sources}

As C24 note, the apparent discrepancy between the inferred luminosity and the accretor's Eddington luminosity 
is already seen in
the study of ultraluminous X-ray sources (ULXs), and so we briefly discuss these systems here. 

The study of ULXs dates back at least to 
\citet{Fabbiano0189}, who gave a list of 16 off-galactic-centre X-ray sources with luminosities of $> 10^{39}\ergs$, 
and there are now more than 1800 known systems. There have been a number of models suggested for ULXs (see \citealt{king0623} for a review), but the situation is now much clearer \citep{Lasota1223}. 

Very recently, polarimetric X-ray observations (\citealt{Veledina0323}) of the X-ray binary Cyg X-3 decisively showed that the behaviour of  ULXs results from geometric beaming (more precisely,  collimation) of outgoing emission towards the observer, as long suspected (\citealt{King0501}). In Cyg X-3, the beam robustly inferred from the polarisation measurements has an angular width of $4\pi b$ steradians, with $b \simeq 1/65$, but lies across the line of sight. {As a result,} the beamed energy is not seen directly; that is, Cyg X-3 is not a ULX but a ULX seen side-on. There are strong arguments (\citealt{Begelman0706,Middleton0921}) that the extreme Galactic system SS 433 is another example of this type. 

ULX beaming occurs because a potentially super--Eddington mass supply rate is
systematically expelled from each radius of the accretion disc by radiation pressure and is progressively reduced to the Eddington value near the accretor. 
As the accretion energy release at disc radius $R$ is proportional to $GM\dot M(R)/R$, this means that the accretion rate within the disc decreases as $\dot M(R) \propto R$.
The total power of the emitted radiation is then $\simeq L_{\rm Edd}(1 + \ln(\dot M/\dot M_{\rm Edd}))$, 
where $\dot M > \dot M_{\rm Edd}$ is the mass supply rate to the disc from outside (\citealt{Shakura73}), and $\dot M_{\rm Edd} = L_{\rm Edd}/\eta c^2$, 
where $\eta \simeq 0.1$ is the accretion efficiency
(see \citealt{KingSMBH} and \citealt{king0623} for discussions). 

The radiation-pressure-driven outflow from the disc is largely optically thick, except that centrifugal repulsion leaves twin narrow vacuum channels
along the accretion disc axis. A luminosity 
$\simeq L_{\rm Edd}$ emitted by the disc eventually finds these channels and escapes to infinity in a narrow double beam \citep{King0902}, while a luminosity of a similar order is emitted from the outer photosphere of the wind (cf \citealt{King0116}) but spread
over $\sim 4\pi$ steradians.

In sources directly identified as ULXs ---that is, with apparently highly super-Eddington luminosities---, terrestrial observers lie within the beam. The assumption of isotropic emission then strongly overestimates the total radiation output, as the specific intensity is far higher within the two beams than in the $\sim 4\pi$ steradians outside them. This means that most X-ray binaries with super-Eddington mass-transfer rates do not directly appear as ULXs, as the observer has to be favourably located.

The beaming factor $b$ {giving the ratio of
the true luminosity $\simeq L_{\rm Edd}$ to the 
apparent (assumed isotropic) luminosity 
$L_{\rm sph}$ of a ULX}
(so that $L_{\rm sph} \simeq (1/b)L_{\rm Edd}$) is given by
\begin{equation}
b \simeq \frac{73}{\dot m^2},
\label{b}
\end{equation}
\citep{King0902}. Here, $\dot m \gg 1$ is the ratio of the mass-supply rate at the outer edge of the accretion disc to the value $\dot M_{\rm Edd} \simeq 10^{-20}L_{\rm Edd}$ that would produce the Eddington luminosity. As this process generally involves only electron scattering, the form of Eq. (\ref{b}) is in practice 
universal for all the radiation produced deep in the potential well near the disc centre.

The $b \propto \dot m^{-2}$  dependence of Eq. (\ref{b})
arises because conditions close to the accreting compact object always asymptote to
the same  Eddington flow near the photosphere, regardless of how potentially super-Eddington the mass supply at the outer disc edge is. This same reasoning (King, 2009) also predicts a relation 
\begin{equation}
L \propto T^{-4}
\end{equation}
for the luminosity $L_{\rm sph}$ and effective temperature $T$ of beamed blackbody emission. This is observed
for `ultrasoft' ULXs, in the form
\begin{equation}
    L_{41} \simeq T_6^{-4}
    \label{LT}
\end{equation}
(\citealt{Kajava0909}), where $L_{41}, T_6$ are $L_{\rm sph}, T$ in units of $10^{41}\ergs$ and $10^6$~K, respectively. {The normalisation of this relation fixes the proportionality factor 73 in Eq. (\ref{b}).} 
We show in Section 3 that the ultrasoft component dominates the observed radiation output of SN 2022jli.

\section{SN 2022jli and ULXs}
We can now see how SN 2022jli can appear with an apparently super-Eddington luminosity of $L_{\rm sph} \simeq 10^{42}\ergs$. 
Comparing with Eq. (\ref{ledd}) and using Eq. (\ref{b}) gives
\begin{equation}
\dot m = \frac{270}{m_{10}^{1/2}},
\label{dotm}
\end{equation}
{which ensures that $b \simeq 10^{-3}$.}

{Equation (\ref{dotm})  tells us that the companion star currently transfers mass towards the compact accretor at a rate
\begin{equation}
    \dot M_{\rm trans} \simeq 2.5\times 10^{-5}m_{10}^{1/2}~\Msun {\rm yr}^{-1}.
    \label{trans}
\end{equation}
This very high rate arises because of the violent rearrangement of the binary geometry caused by the SN, which has evidently forced the companion star to overflow its Roche lobe significantly.
All but a small part ($\simeq \dot M_{\rm Edd} \sim 10^{-7}m_{10}~\Msun {\rm yr}^{-1}$) of this transfer rate is ultimately ejected to infinity. The SN is also probably responsible for the system's relatively long current orbital period, as it almost succeeded in unbinding the binary. }

An important feature of SN 2022jli is that almost all of its luminosity is emitted in the UVOIR region of the electromagnetic spectrum, rather than in X-rays as one might initially expect. This occurs in ultrasoft ULXs where the mass-transfer rate is particularly high (\citealt{Kajava0909}),
making the optical depth through the outflow very large: all the photons that emerge have undergone many slightly inelastic scatterings. Because of its very soft spectrum, we classify SN 2022jli as an ultrasoft ULX.
We see from Eq. (\ref{LT}) that this very soft spectrum is as expected for the inferred value of the apparent luminosity of $L_{\rm sph} \simeq 10^{42}\ergs$, which gives $T \sim 5\times 10^5$~K{\footnote{We note that although Eq. (\ref{LT}) formally predicts that the blackbody emission would eventually appear in medium-energy X-rays
-- i.e. with $T_6 = 10$ -- during the final fast luminosity decay, but the luminosity
would then only be $\sim 10^{37}\ergs$, making this effectively unobservable.}}. We can use this value to check that the opacity of the gas outflow is indeed dominated by electron scattering.\footnote{We stress that Eq. \ref{LT} applies only to sources being supplied with mass at a strongly super-Eddington rate. The peak X-ray luminosity of the famous hyperluminous source HLX-1 in the galaxy ESO 243–49 is $> 10^{42}\ergs$ \citep{Farrell0709}, but the mass of its black hole is presumably $ >10^4\Msun$, and so it is fed mass at a near-Eddington rate \citep{Godet0617}. 
This system may be a long-period quasi-periodic eruption (QPE) source (\citealt{King0922,Webb0523}), where a star (in practice a white dwarf) that has narrowly avoided complete tidal disruption by the black hole periodically fills its tidal lobe at the pericentre of an extremely eccentric orbit about the black hole \citep{Godet1014,King0922}.}

The sudden loss of mass from the star that exploded as a SN, leaving only a compact remnant accretor, must make the binary extremely eccentric; indeed it is only the asymmetric kick which holds the binary together, keeping the eccentricity $e$ below unity. The high eccentricity is also the reason why the binary transfers mass at all, as otherwise the effective Roche lobe (see Eq. (\ref{lobe}) below)
is too large for the companion star to fill it. We demonstrate below that for a solar-type star to fill the lobe, we need $1 - e \lesssim 0.045$ or $e \gtrsim 0.95$. 

For such extreme eccentricities (actually $e \gtrsim 0.97$, \citealt{King0723,Miniutti0223}), the viscous timescale in the accretion disc is shorter than the binary period of $P = 12.4$~days. The observed light curve therefore reflects the periodic bursts of mass transfer as the companion star passes through the binary pericentre:
each burst is largely accreted by the black hole or neutron star before the next occurs 12.4 days later.

The stability of the light curve also suggests that the accretor (black hole or neutron star) is relatively fixed near the centre of mass of the system, with its radiation beam oriented stably towards us. This must mean that the companion star has a significantly lower mass than the accretor. SN2022jli therefore represents the birth event of a low-mass X-ray binary (LMXB).

The asymmetric nature of the SN kick also makes it very likely that the spin axis of the accreting black hole or neutron star is misaligned with respect to  the binary orbit. Viscous torques caused by the resulting precession (e.g Lense--Thirring) then cause the central accretion disc to be warped into the spin plane. Since SN2022jli appears as a ULX, this plane is orthogonal to the line of sight along the vacuum funnels through the outflow. The binary orbital plane is misaligned with respect to this axis, and so we observe periodic Doppler shifts resulting from its orbital motion.

\section{The evolution of SN 2022jli}

The new eccentric binary LMXB that SN 2022jli created must evolve very rapidly because of its high mass-transfer rate (\ref{trans}) caused by the {violent} effect of the SN explosion. This is far larger than effects such as orbital angular momentum loss, and implies a typical current evolution timescale of 
\begin{equation}
t_{\rm evol} \sim \frac{M_2}{-\dot M_2} \sim 10^5\,{\rm yr},
\label{evol}
\end{equation}
where $M_2 \sim 1 \Msun$ is the companion mass. 
The binary angular momentum $J$ is 
\begin{equation}
    J = M_1M_2\left(\frac{Ga}{M}\right)^{1/2}(1 - e^2)^{1/2}
    \label{J}
,\end{equation}
where $M_1$ is the compact object mass, $M = M_1+M_2$ is the total binary mass, and $a$ is the binary separation. During the evolution, we have $\dot M_2 <0,\dot M_1 \simeq 0,$ and $\dot M = \dot M_2$. In the current phase, we can neglect systemic angular momentum loss and eccentricity evolution, and so logarithmic differentiation of (\ref{J}) gives
\begin{equation}
\frac{\dot a}{a} \simeq - \frac{2\dot M_2}{M_2} > 0.
\label{delta a}
\end{equation}
The binary expands because mass is being transferred closer to the centre of mass while conserving total angular momentum. 
This expansion must significantly reduce the mass-transfer rate and therefore also the accretion luminosity once the increase in separation 
$\Delta a$ becomes of order the atmospheric scale height of the companion star
\footnote{We note that a main sequence star does not expand on adiabatic mass loss unless its mass is low, i.e. $\lesssim 0.3\Msun$.}.
This is 
\begin{equation}
    H = \frac{kT_2R_2^2}{GM_2\mu m_H} \sim 5\times 10^7~{\rm cm},
\label{H}
\end{equation}
where we have taken solar values for the surface temperature $T_2$ and radius $R_2$ ($\mu$ and $m_H$ are respectively the mean molecular and hydrogen mass). For a star filling the modified Roche lobe for a binary of eccentricity $e,$ we have
\begin{equation}
    \frac{R_2}{a} \simeq 0.46\left(\frac{M_2}{M_1}\right)^{1/3}(1 - e),
    \label{lobe}
\end{equation}
and again assuming $M_2 \sim 1\Msun, M_1 \sim 10M_2$, and $R_2 \sim R_{\odot}$, we find 
$a \simeq 3\times 10^{12}$~cm. Comparing with Eq. (\ref{H}) gives $H\lesssim 10^{-5}a$. From 
Eq. (\ref{delta a}) in the form of $\Delta a/a \sim -2\Delta M_2/M_2$,
this shows that the initial mass-transfer rate of the binary must decrease on a 
timescale of $\lesssim 1$~yr as $\Delta a \sim H$. 
This is evidently the origin of the observed luminosity decay over $\sim 250$~days. 

This is a very rare case where the very high mass-transfer rate (caused here by the SN explosion) makes the Roche lobe move through a stellar scale height on a timescale so short that it is observable. This means that the observed mass-transfer rate is close to the (changing) evolutionary mean. In general, this condition does not hold, and observed changes in mass-transfer rates are completely unrelated to the evolutionary mean (see \citealt{King1221} and references therein).

\section{Evolution to a low-mass X-ray binary}

This rapid decay of the impulsive mass transfer resulting from the destabilising effect of the SN explosion will soon cause the system to detach.
This
opens various possible routes to the system's probable endpoint as a 
LMXB. 

Systemic angular momentum loss (AML) will tend to reverse the current decaying orbital expansion. However, at the current (and lengthening) 12.4 d binary period, the gravitational radiation timescale exceeds a Hubble time, even given the likely high orbital eccentricity{, and so this form of AML is unlikely to restart mass transfer. The other possible AML process is magnetic stellar wind braking of the companion star spin, which is transmitted to the orbit via tides. Current treatments
only consider the case of circular binaries, and so again estimates are problematic.  

Regardless of these AML effects, a likely upper limit to the timescale for reaching the LMXB state is the main sequence lifetime of the companion star. 
If AML has not yet shrunk the binary and restarted mass transfer by the time that the companion leaves the main sequence, the nuclear evolution of the companion will drive mass transfer instead, and the system will become a relatively long-period (days -- years) LMXB.

\section{A possible observational test}

The ideas of this paper are in principle open to an observational test. From Kepler's law, the expansion of the system driven by the current very rapid mass transfer implies a rate of period increase of 
\begin{equation}
    \frac{\dot P}{P} = \frac{3\dot a}{2a} - \frac{\dot M}{2M} \sim 10^{-5}\, {\rm yr}^{-1},
    \label{dotp1}
\end{equation}
which implies
\begin{equation}
\dot P \sim 3\times 10^{-7} {\rm s\, s}^{-1}.
\label{dotp2}
\end{equation}
This gives a direct test of the deduced mass-transfer value (\ref{trans}), but may be difficult to measure as the system fades on the current short timescale.

\section {Conclusion}

Most of the distinctive features of SN 2022jli come from the violent disturbance to the binary geometry caused by the sudden mass loss in the SN explosion. The ultraluminous nature of the current accretion onto the newborn compact component allows us to quantify the resulting mass-transfer rate and its effects on the binary. Unusually, these include changes in the binary separation comparable to the density scale height of the donor star on observable timescales, with consequent effects on the mass-transfer rate, as detailed in Section 6.

\section*{Acknowledgments}
{We thank the anonymous referee and the Editor of the paper for perceptive and helpful comments.}
\bibliographystyle{aa}
\bibliography{ULS} 
\end{document}